\documentclass[prl,twocolumn,showpacs]{revtex4}
\usepackage{graphicx}
\usepackage{float}
\usepackage{amsmath}
\usepackage{amssymb}

\begin{document}

\title{Transformation Optics Approach to Plasmon-Exciton Strong Coupling in Nanocavities}
\author{Rui-Qi Li$^{1,2}$, D. Hern\'angomez-P\'erez$^{1}$, F. J. Garc\'ia-Vidal$^{1,3}$, and A. I.
Fern\'andez-Dom\'inguez$^{1}$}

\affiliation{$^{1}$Departamento de F\'isica Te\'orica de la
Materia Condensada and Condensed Matter Physics Center (IFIMAC),
Universidad Aut\'onoma de Madrid, E- 28049 Madrid, Spain,
$^{2}$Key Laboratory of Modern Acoustics, MOE, Institute of
Acoustics, Department of Physics, Nanjing University, Nanjing
210093, People's Republic of China, $^{3}$Donostia International
Physics Center (DIPC), E-20018 Donostia/San Sebasti\'an, Spain}

\begin{abstract}
We investigate the conditions yielding plasmon-exciton strong
coupling at the single emitter level in the gap between two metal
nanoparticles. A quasi-analytical transformation optics approach
is developed that makes possible a thorough exploration of this
hybrid system incorporating the full richness of its plasmonic
spectrum. This allows us to reveal that by placing the emitter
away from the cavity center, its coupling to multipolar dark modes
of both even and odd parity increases remarkably. This way,
reversible dynamics in the population of the quantum emitter takes
place in feasible implementations of this archetypal nanocavity.\
\end{abstract}


\maketitle

Plasmon-exciton-polaritons (PEPs) are hybrid light-matter states
that emerge from the electromagnetic (EM) interaction between
surface plasmons (SPs) and nearby quantum emitters
(QEs)~\cite{Chang2006,Torma2015}. Crucially, PEPs only exist when
these two subsystems are strongly coupled, i.e., they exchange EM
energy coherently in a time scale much shorter than their
characteristic lifetimes. Recently, much attention has focused on
PEPs, since they combine the exceptional light concentration
ability of SPs with the extreme optical nonlinearity of QEs. These
two attributes makes them promising platforms for the next
generation of quantum nanophotonic components~\cite{Tame2013}.

A quantum electrodynamics description of plasmonic strong coupling
of a single QE has been developed for a flat metal
surface~\cite{GonzalezTudela2014}, and
isolated~\cite{Trugler2008,Waks2010} and distant
nanoparticles~\cite{Savasta2010,Manjavacas2011,Esteban2014}, where
SP hybridization is not fully exploited. From the experimental
side, in recent years, PEPs have been reported in emitter
ensembles~\cite{Bellesa2004,Schwartz2011,Zengin2015,Todisco2015},
in which excitonic nonlinearities are negligible
~\cite{Salomon2012,GonzalezTudela2013,Delga2014}. Only very
recently, thanks to advances in the fabrication and
characterization of large Purcell enhancement
nanocavities~\cite{Hartsfield2015,Gong2015,Hoang2016}, far-field
signatures of plasmon-exciton strong coupling for single molecules
have been reported experimentally~\cite{Chikkaraddy2016}.

\begin{figure}[!ht]
\begin{center}
\includegraphics[angle=0,width=0.48\textwidth]{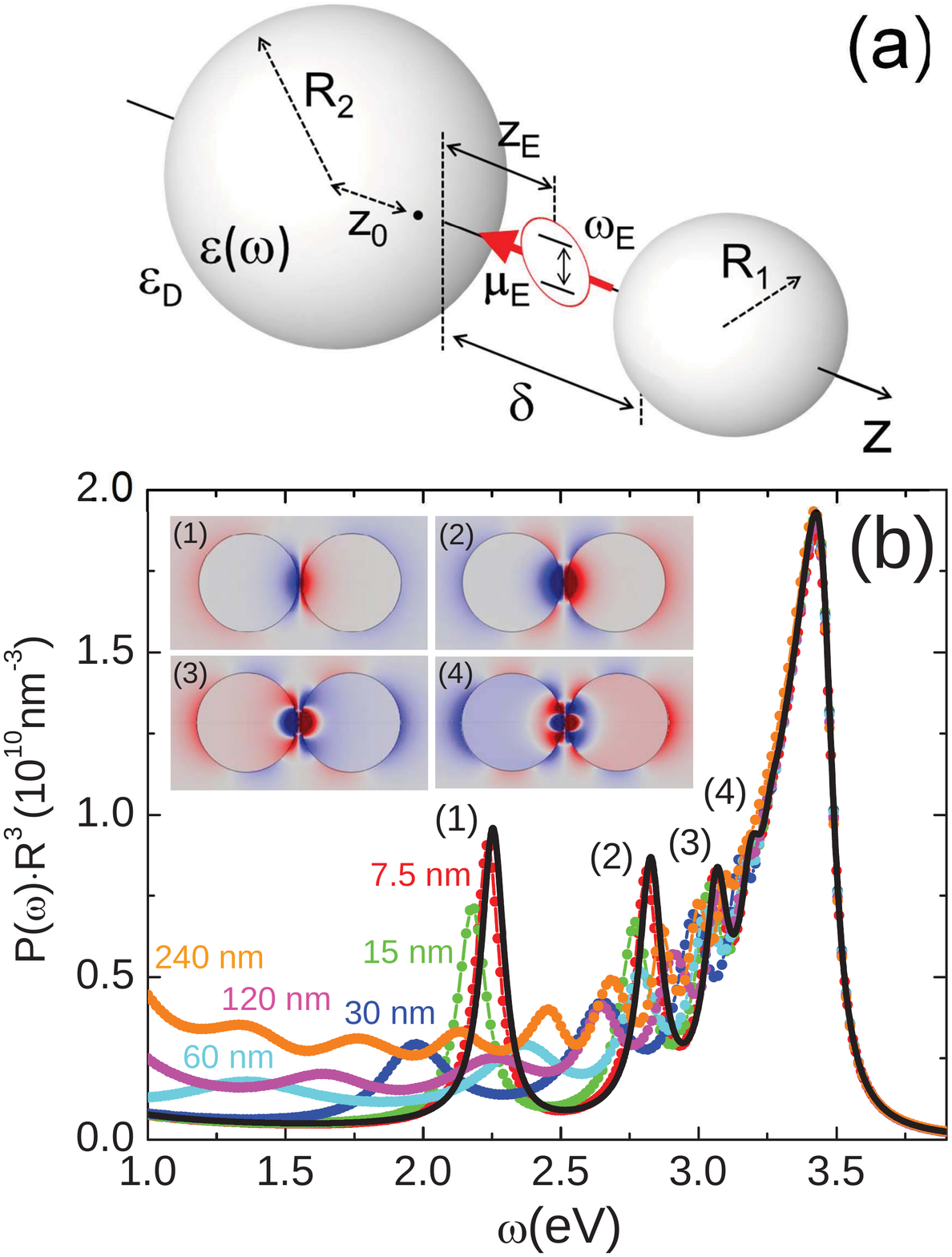}
\end{center}
\vspace{-0.7cm}\caption{(a) QE placed at the gap between two metal
spheres of permittivity $\epsilon(\omega)$ and embedded in a
dielectric medium $\epsilon_{\rm D}$. The QE dipole strength,
position and frequency are $\mu_{\rm E}$, $z_{\rm E}$ and
$\omega_{\rm E}$. (b) Normalized Purcell factor at the gap center
for $R_{1,2}=R$ and $\delta=R/15$. Color dots: EM simulations for
different $R$. Black line: TO prediction. Insets: Induced charge
distribution for the lowest 4 SP modes discernible in the TO
spectrum (color scale is saturated for clarity).} \vspace{-0.5cm}
\end{figure}

In this Letter, we theoretically investigate the plasmonic
coupling of a single emitter in a paradigmatic cavity: the
nanometric gap between two metallic
particles~\cite{Zengin2015,Hoang2016,Chikkaraddy2016}. We consider
spherical-shaped nanoparticles, and develop a transformation
optics (TO)~\cite{Pendry2012,Pendry2013} approach that fully
accounts for the rich EM spectrum that originates from SP
hybridization across the gap. Our method, which is the first
application of TO concepts to treat quantum optical phenomena,
yields quasi-analytical insight into the Wigner-Weisskopf
problem~\cite{Petruccionebook} for these systems, and enables us
to reveal the prescriptions that nanocavities must fulfil to
support single QE PEPs.

Figure 1(a) sketches the system under study: a two level system
(with transition frequency $\omega_{\rm E}$ and $z$-oriented
dipole moment $\mu_{\rm E}$) placed at position $z_{\rm E}$ within
the gap $\delta$ between two spheres of permittivity
$\epsilon(\omega)=\epsilon_\infty-\tfrac{\omega_{\rm
p}^2}{\omega(\omega+i\gamma)}$, embedded in a matrix of dielectric
constant $\epsilon_{\rm D}$ [see Supplemental Material (SM) for
further details]. We assume that the structure is much smaller
than the emission wavelength and operate within the quasi-static
approximation. The details of our TO description of SP-QE coupling
in this geometry, based on the method of
inversion~\cite{Landaubook,Pendry2013}, can be found in the SM.
Briefly, by inverting the structure with respect to a judiciously
chosen point [$z_0$ in Figure 1(a)], the spheres map into an
annulus geometry in which the QE source and scattered EM fields
are expanded in terms of the angular momentum $l$. This allows us
to obtain the scattering Green's function, $G^{\rm
sc}_{zz}(\omega)$, in a quasi-analytical fashion.

First we test our TO approach by analyzing the spontaneous
emission enhancement experienced by an emitter at the gap center.
Figure 1(b) plots the Purcell factor $P(\omega)=1+\frac{6\pi
c}{\omega} {\rm Im}\{G^{\rm sc}_{zz}(\omega)\}$ for dimers with
$R_{1,2}=R$. To compare different sizes, $P(\omega)$ is normalized
to $R^{-3}$. Black solid line plots the TO prediction (identical
for all sizes), and color dots render full EM calculations (Comsol
Multiphysics$^{\rm TM}$). At high frequencies, TO and simulations
are in excellent agreement for all $R$. At low frequencies,
discrepancies caused by radiation effects are evident for
$R\gtrsim 30$ nm. The insets in Figure 1(b) render induced charge
density maps for the four lowest peaks in the TO spectrum. These
can be identified as SP resonances of increasing multipolar order.
We can infer that the maximum that dominates all the spectra in
Figure 1(b) is caused by the pseudomode ($\omega_{\rm PS}$)
emerging from the spectral overlapping of higher order
SPs~\cite{Delga2014}. Importantly, these are darker (weakly
radiative) modes strongly confined at the gap region, which
explains why quasi-static TO is valid at $\omega_{\rm PS}$ even
for $R=240$ nm.

\begin{figure}[!t]
\begin{center}
\includegraphics[angle=0,width=0.47\textwidth]{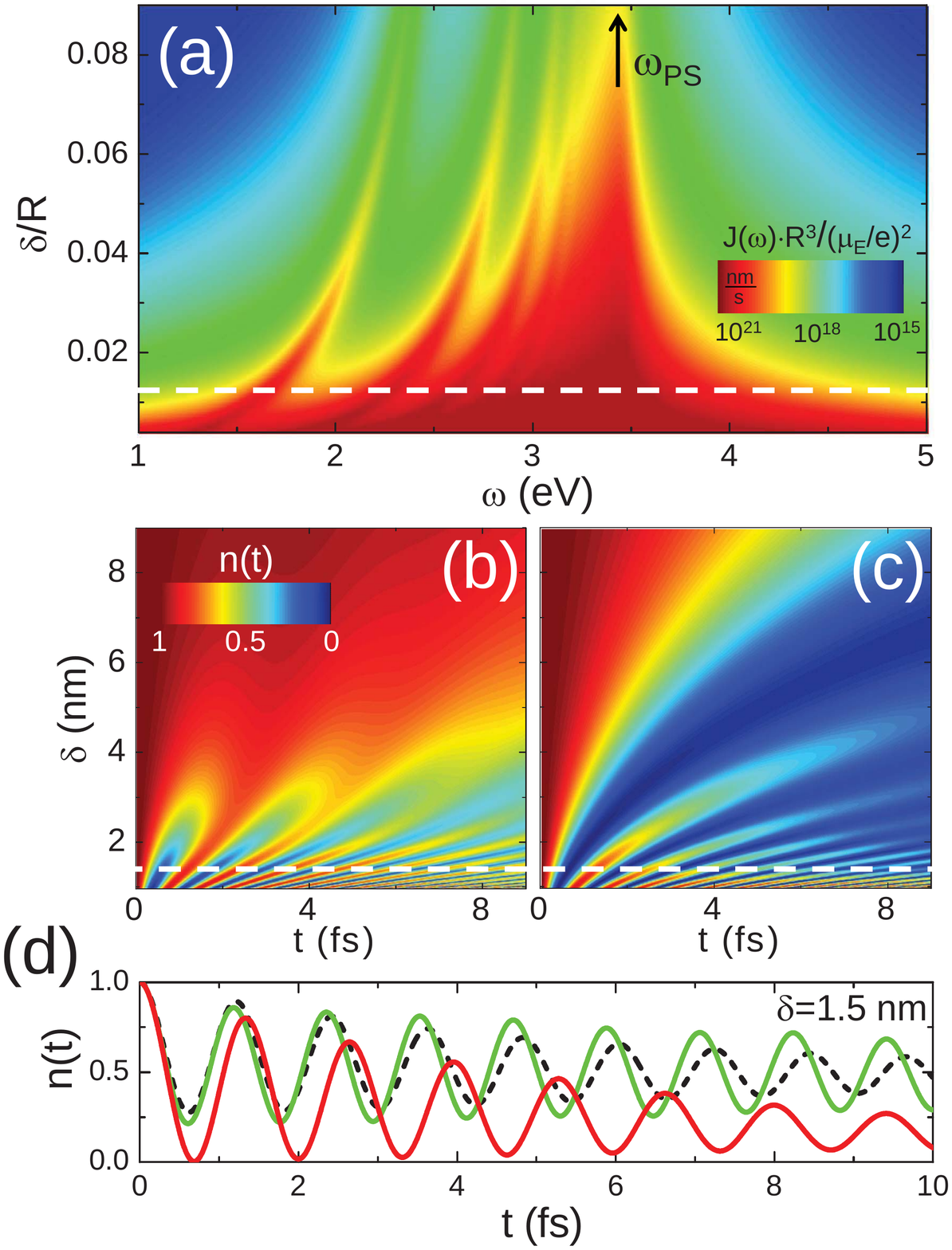}
\end{center}
\vspace{-0.7cm} \caption{(a) Normalized $J(\omega)$ at the gap
center versus frequency and $\delta/R$. (b-c) $n(t)$ versus time
and gap size for $R=120$ nm and $\mu_{\rm E}=1.5 \ {\rm
e}\cdot\textrm{nm}$. The QE is at resonance with the dipolar SP
mode in (b) and with the pseudomode in (c). (d) $n(t)$ for
$\delta=1.5$ nm (see white dashed lines) and two $\omega_{\rm E}$:
1.7 (green) and 3.4 (red) eV. Black dotted line corresponds to
$\omega_{\rm E}=1.7$ eV obtained through the fitting of
$J(\omega)$ at $\omega_{\rm PS}$.}
\end{figure}

Now we investigate the spectral density across the gap cavity.
This magnitude governs SP-QE interactions (see below), and can be
expressed as $J(\omega)=\tfrac{\mu_{\rm
E}^2\omega^3}{6\pi^2\epsilon_0\hbar c^3}P(\omega)$. Figure 2(a)
shows TO-$J(\omega)$ evaluated at $z_{\rm E}=\delta/2$ and
normalized to $\mu_{\rm E}^2/R^3$ for different $\delta/R$. For
small gaps, the spectral density is maximized, and the
contribution from different SPs is apparent. For larger gaps,
$J(\omega)$ decreases, all maxima blue-shift and eventually merge
at the pseudomode position. Importantly, Figure 2(a) shows a
universal trend, valid for all QEs and $R$ (within the
quasi-static approximation). Therefore, for a given $\delta/R$,
large $\mu_{\rm E}$ and small $R$ must be used to increase
plasmon-exciton coupling.

Once the spectral density is known, the Wigner-Weisskopf
problem~\cite{Petruccionebook} can be solved. It establishes that
the equation governing the dynamics of the excited-state
population, $n(t)=|c(t)|^2$, for an initially excited QE is
\begin{equation}
\frac{d}{dt} c(t)=-\int_0^t d\tau \int_0^\infty d\omega
J(\omega)e^{i(\omega_{\rm E}-\omega)(t-\tau)}c(\tau).
\end{equation}
Figure 2(b-c) render the QE population at the center of the cavity
in panel (a) as a function of time and gap size. The spheres
radius is 120 nm (so that $1\lesssim\delta\lesssim 10$~nm), and
$\mu_{\rm E}=1.5 \ {\rm e}\cdot{\rm nm}$ (InGaN/GaN quantum dots
at 3 eV~\cite{Ostapenko2010}). The emitter is at resonance with
the lowest (dipolar) SP (b) and with the pseudomode (c) maxima in
Figure 2(a), respectively. Note that the former disperses with gap
size, whereas $\omega_{\rm E}=\omega_{\rm PS}$ for the latter. We
can observe that both configurations show clear oscillations in
$n(t)$, which indicates that coherent energy exchange is taking
place. In this regime, strong coupling occurs, and the nanocavity
supports PEPs. However, for $\delta> 3$ nm, the reversible
dynamics in the population is lost in both panels, QEs and SPs are
only weakly coupled, and $n(t)$ follows a monotonic decay.

Figure 2(d) plots $n(t)$ at strong coupling, $\delta=R/80=1.5$ nm
[see white dashed lines in panels (a-c)]. The red (green) line
corresponds to QE at resonance with the pseudomode (dipolar SP)
peak. The excited state population obtained from the fitting of
$J(\omega)$ around $\omega_{\rm PS}$ and evaluated at the lowest
SP frequency is shown in black dashed line. The similarity between
solid green and dashed black lines implies that the population
dynamics is fully governed by the pseudomode, even when the two
maxima in $J(\omega)$ are far apart (the differences between
Figures 2(b) and (c) originate from detuning effects). This fact
enables us to extend the validity of our TO approach to larger
structures, as radiative effects do not play a significant role at
the pseudomode. More importantly, our findings reveal that QE
strong coupling in nanocavities does not benefit from highly
radiative plasmonic modes despite their low resonant frequencies
and associated low sensitivity to metal absorption.

\begin{figure}[!t]
\begin{center}
\includegraphics[angle=0,width=0.45\textwidth ]{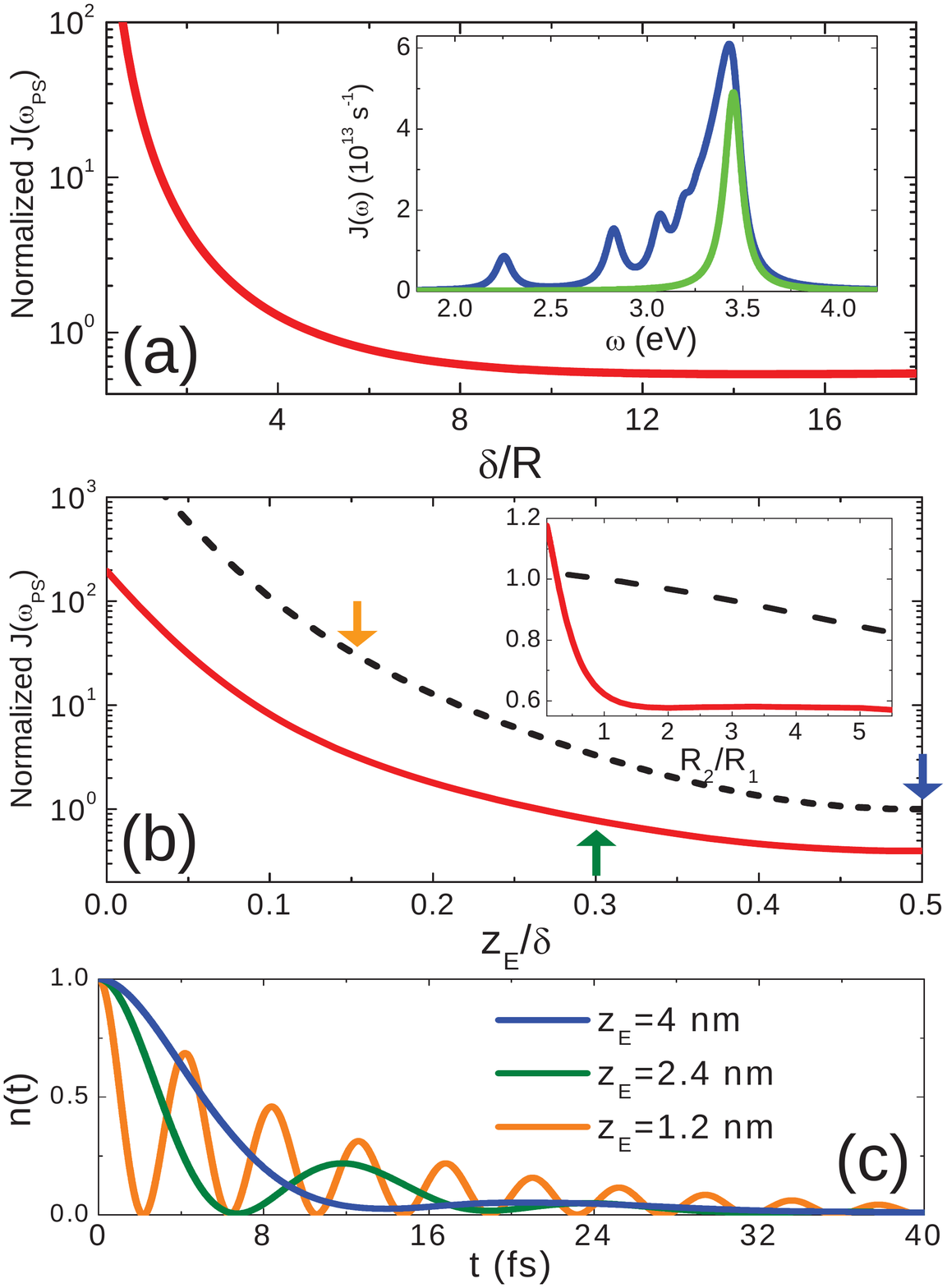}
\end{center}
\vspace{-0.9cm}\caption{(a) $J(\omega)$ at $z_E=\delta/2$ and
$\omega_{\rm E}=\omega_{\rm PS}$ versus $\delta$ normalized to the
sum of the spectral density maxima for the spheres isolated.
Inset: $J(\omega)$ for the dimer (blue) and isolated particle
(green) for $\delta=8$ nm, R=120 nm. (b) Spectral density at the
pseudomode versus $z_{\rm E}/\delta$. Red solid line:
$J(\omega_{\rm PS})$ normalized to the sum of the two spheres
isolated. Black dashed line: $J(\omega_{\rm PS})$ normalized to
its value at $z_{\rm E}=\delta/2$. Inset: Same but versus the
ratio $R_2/R_1$ for $z_{\rm E}=\delta/2$. (c) $n(t)$ for
$\omega_{\rm E}=\omega_{\rm PS}$ and three $z_{\rm E}$ values
($\mu_{\rm E}=1.5 \ {\rm e}\cdot{\rm nm}$).}
\end{figure}

We have found that $R=120$ nm cavities can support single QE PEPs
only if $\delta<4$ nm. Similar calculations for single particles
(not shown here) indicate that the onset of strong coupling takes
place at similar distances, $z_{\rm E}\lesssim 2$ nm. This means
that the configuration investigated so far does not exploit
cooperative effects between the nanospheres, associated to the
enhancement in $J(\omega)$ expected from SP hybridization. To
verify this, Figure 3(a) plots $J(\omega_{\rm PS})$ versus
$\delta$ evaluated at the center of the cavity and normalized to
twice the maximum in the spectral density for an isolated sphere
($R=120$ nm, $z_{\rm E}=\delta/2$). Whereas normalized
$J(\omega_{\rm PS})$ is much larger than 1 for $\delta=1.5$ nm, it
decays to $\sim 0.5$ for gaps larger than 4 nm. Therefore, only
very small gap cavities take advantage of SP hybridization. The
inset of Figure 3(a) plots $J(\omega)$ for 120 nm radius dimer
(blue) and single sphere (green) evaluated at $z_{\rm E}=4$ nm,
showing that the maximum spectral density is very similar in both
cases.

We explore next the effect that moving the QE away from the gap
center has on the cavity performance. We consider $\delta=8$ nm,
for which strong coupling does not take place at $z_{\rm
E}=\delta/2$, see Figure 2(b-c). Figure 3(b) plots $J(\omega_{\rm
PS})$ versus $z_{\rm E}$ for two different normalizations. Black
dashed line shows the ratio of $J(\omega_{\rm PS})$ and its value
at $z_{\rm E}=\delta/2$. We can observe that the spectral density
maximum grows exponentially as the QE approaches one of the
particles, yielding factors up to $10^3$. This effect could be
attributed to the stronger interaction with the SPs supported by
the closest sphere. To test this, red solid line plots
$J(\omega_{\rm PS})$ now normalized to the sum of the spectral
densities calculated for each of the spheres isolated and
evaluated at $z_{\rm E}$ and $\delta-z_{\rm E}$. Remarkably,
enhancements up to $10^2$ are found in this asymmetric
configuration. Therefore, the pronounced increase of $J(\omega)$
cannot be simply caused by proximity effects, but it must be due
to a significant enhancement of the cooperativity between the two
nanoparticles. Figure 3(c) plots $n(t)$ for three $z_{\rm E}$
values (indicated by vertical arrows in panel (b)), proving that
strong coupling occurs for $z_{\rm E}$ far from the cavity center.
The inset of Figure 3(b) investigates if SP-QE coupling can
benefit further from geometric asymmetry. It renders
$J(\omega_{\rm PS})$ versus $R_2/R_1$ for both normalizations, and
proves that the cavity performance is rather independent of the
particle sizes in the regime $R_{1,2}\gg\delta$.

To gain physical insight into the dependence of $J(\omega)$ on the
QE position, we assume that $\delta\ll R_{1,2}$, and work within
the high quality resonator limit~\cite{Waks2010}. This way, we can
obtain analytical expressions for $J(\omega)$, which can be
written as a sum of Lorentzian SP contributions of the form
\begin{equation}
J(\omega)=\sum_{l=0}^\infty\sum_{\sigma=\pm
1}\frac{g_{l,\sigma}^2}{\pi}\frac{\gamma/2}{(\omega-\omega_{l,\sigma})^2+(\gamma/2)^2},
\end{equation}
where the index $l$ can be linked to the multipolar order of the
SP, $\sigma$ to its even (+1) or odd (-1) character, and $\gamma$
is the damping parameter in $\epsilon(\omega)$.

The SP resonant frequencies in Equation~(2) have the form
\begin{equation}
\omega_{l,\sigma}=\frac{\omega_{\rm
p}}{\sqrt{\epsilon_\infty+\epsilon_{\rm
D}\tfrac{\xi_l+\sigma}{\xi_l-\sigma}}},
\end{equation}
with
$\xi_l=\big[\frac{(3R+\delta-z_0)(R+\delta-z_0)}{(R-z_0)(R-z_0)}\big]^{l+\tfrac{1}{2}}$.
Note that, for simplicity, we focus here in the case $R_{1,2}=R$,
but general expressions can be found in the SM. Importantly, for
large $l$, $\xi_l\gg 1$, which enables us to write $\omega_{\rm
PS}\sim\tfrac{\omega_{\rm p}}{\sqrt{\epsilon_\infty+\epsilon_{\rm
D}}}$. The spectral overlapping giving rise to the pseudomode
always peaks at a frequency slightly lower than the SP asymptotic
frequency for a flat metal surface.

The coupling constants, $g_{l,\sigma}$, in Equation~(2) are
mathematically involved functions of the geometric parameters of
the cavity. However, without loss of generality, we can write
\begin{equation}
g_{l,\sigma}^2=\frac{\mu_{\rm
E}^2}{\Delta^3}f\Big(\tfrac{\Delta}{z_{\rm E}+R-z_0}\Big),
\end{equation}
where $f(\cdot)$ contains all the dependence on the emitter
position and
$\Delta=\frac{(R+\delta-z_0)(3R+\delta-z_0)}{2R+\delta-z_0}$ gives
the inverse volume scaling of $J(\omega)$ anticipated in Figure 1.
Equation (4) proves formally that the cavity performance can be
improved by reducing its overall size, as this increases the
coupling strength for all SP modes. Let us remark that the
analytical decomposition of $J(\omega)$ given by Equations (2)-(4)
proves the suitability of TO for the description of quantum
nano-optical phenomena. It provides naturally a convenient and
efficient quantization of EM fields in lossy, complex
nanocavities, a research area of much theoretical activity
lately~\cite{Doost2014,Kristensen2014}.

In the following, we test our analytical approach. Figure 4(a)
plots $J(\omega)$ for the case $z_{\rm E}=0.3\delta$ in Figure
3(c). Red dashed-dotted and black dashed lines plot exact TO and
EM calculations, respectively. The spectrum obtained from
Equation~(2) is rendered in green solid line. It reproduces
$J(\omega)$ satisfactorily except for a small red-shift in the
lowest frequency peak (with respect to the exact TO prediction).
The various contributions to $J(\omega)$ in Equation~(2) are
plotted in blue dashed and solid orange lines in Figure 3(a).
These two sets correspond to even ($\sigma=+1$) and odd
($\sigma=-1$) SP modes, respectively. Note that the former
(latter) blue-shift (red-shift) towards $\omega_{\rm PS}$ for
increasing $l$. These different trends originate from the ratio
$\tfrac{\xi_l+\sigma}{\xi_l-\sigma}$ in the denominator of
Equation~(4), which is always larger (smaller) than 1 for
$\sigma=+1$ ($\sigma=-1$). The insets of Figure 4(a) depict
induced surface charge density maps for the maxima corresponding
to the two lowest odd SP contributions. Note that due to their
antisymmetric character, these are purely dark, dipole-inactive,
modes in the quasi-static limit.

\begin{figure}[!t]
\begin{center}
\includegraphics[angle=0,width=0.48\textwidth ]{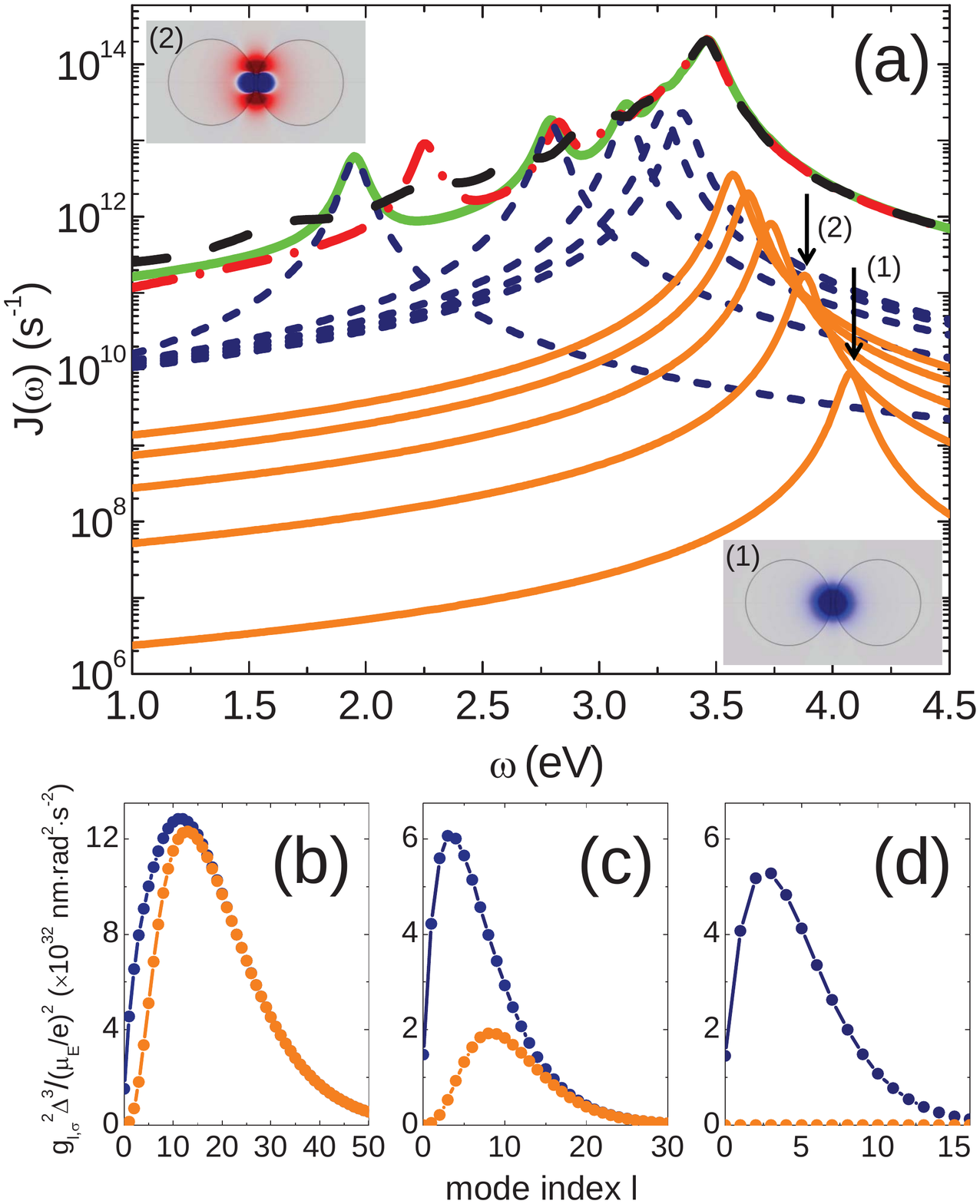}
\end{center}
\vspace{-1.2cm} \caption{(a) Spectral density for $z_{\rm E}=2.4$
nm obtained through numerical (black dashed line), exact TO (red
dotted-dashed line), and analytical TO (solid green line)
calculations. The contribution to $J(\omega)$ due to even and odd
modes are plotted in dark blue dotted and solid orange lines,
respectively. Inset: Surface charge map for the two lowest odd
SPs. (b) Normalized coupling constant squared for even and odd
modes versus $l$ for $z_{\rm E}$: 1.2 nm (b), 2.4 nm (c), and 4 nm
(d).}
\end{figure}

Figures 4(b-d) plot Equation~(4) for both SP symmetries as a
function of the mode index $l$ and evaluated at the three $z_{\rm
E}$'s in Figure 3(c). For QEs in close proximity to one of the
particles ($z_{\rm E}=0.15\delta$), $g_{l,\pm 1}^2$ are largest.
The coupling strength dependence on $l$ is very similar for both
mode symmetries and peaks at $l\simeq 12$. This indicates that
high multipolar dark SPs are responsible for the main
contributions to $J(\omega)$. At intermediate positions, $z_{\rm
E}=0.3\delta$, both coupling constants decrease, being the
reduction much more pronounced in $g_{l,-1}^2$. Finally,
$g_{l,-1}^2$ vanishes at the cavity center ($z_{\rm
E}=0.5\delta$), and the QE interacts only with even SPs having
$l\sim 3$. The bright character of these plasmon resonances
translates into an increase of radiative losses, which worsens
significantly the cavity performance. Figures 4(b-d) evidence that
the remarkable (several orders of magnitude) enhancement in
$J(\omega_{\rm PS})$ shown in Figure 3(b) for $z_{\rm E}$ away
from the $\delta/2$ is caused by two different mechanisms. On the
one hand, the emitter interacts more strongly with even SPs (of
increasing multipolar order). On the other hand, it can couple to
a whole new set of dark modes contributing to $J(\omega)$, those
with odd symmetry, which are completely inaccessible for $z_{\rm
E}=\delta/2$. It is the combination of these two effects which
makes possible to realize plasmon-exciton strong coupling in
nanocavities with $\delta\sim 5-10$ nm.

Finally, in order to prove the predictive value of our TO
analytical method, we calculate the plasmon-exciton coupling
strength for geometrical and material parameters modelling the
experimental samples in Ref.~\onlinecite{Chikkaraddy2016} (see SM
for details). Our approach predicts $g_{0,+1}=19$ meV for the
dipolar SP  mode, and $g^{\rm ef\/f}_{\rm PS}=120$ meV for the
pseudomode. The latter is in good agreement with the measured
value: $g_{\rm exp}=90$ meV. This indicates that, in accordance
with our theoretical findings, high order multipolar dark modes
seem to play a relevant role in the QE-SP interactions taking
place in the nanocavity samples that lead to single molecule
strong coupling.

In conclusion, we have presented a transformation optics
description of plasmon-exciton interactions in nanometric gap
cavities. We have shown that it is the dark pseudomode that builds
up from the spectral overlapping of high frequency plasmonic modes
which governs the energy exchange between emitter and cavity
field. The quasi-analytical character of our approach allows for a
thorough exploration of these hybrid systems, revealing that the
coupling can be greatly enhanced when the emitter is displaced
across the gap. We have obtained analytical expressions that prove
that this increase of the spectral density in asymmetric positions
is caused by not only even, but also odd modes. Finally, we have
verified the predictive value of our analytical approach against
recent experimental data, which demonstrates its validity as a
design tool for nanocavities sustaining plasmon-exciton-polaritons
at the single emitter level.

This work has been funded by the EU Seventh Framework Programme
under Grant Agreement FP7-PEOPLE-2013-CIG-630996, the European
Research Council (ERC-2011-AdG Proposal No. 290981), and the
Spanish MINECO under contracts MAT2014-53432-C5-5-R and
FIS2015-64951-R. R.-Q.L. acknowledges funding by the China
Scholarship Council and thanks Prof. Jian-Chun Cheng for guidance
and support.

\end{document}